\documentclass[final,5p,twocolumn]{elsarticle}
\usepackage{graphicx}
\usepackage{amsmath}
\usepackage{booktabs}
\usepackage{amssymb}
\usepackage{float}
\usepackage{tikz}
\usepackage{caption}
\usepackage{subcaption}
\usepackage{gensymb}
\usepackage{textcomp}
\usepackage{xspace}
\usepackage{siunitx}
\graphicspath{ {./Figures/}
}

\usepackage{hyperref}  % hyper-links 
\usepackage{breakurl}
\hypersetup{breaklinks = true, colorlinks = true, citecolor = blue, linkcolor = blue, urlcolor = blue}

\usepackage{lineno}
\usepackage{lipsum}

\journal{Nuclear Instruments and Methods in Physics Research}

\newcommand{\Moller}{M\o ller\xspace}
\newcommand{\VDG}{Van de Graaff\xspace}
\newcommand{\HVBD}{HV Bias Disc\xspace}
\newcommand\Fig[1]{Fig.~\ref{#1}\xspace} 
\newcommand{\degrees}{\degree\xspace}

\begin{document}

\begin{frontmatter}

\title{Realization of a Large-Acceptance Faraday Cup for 3\,MeV Electrons}

\author[label1]{R.~Johnston\corref{cor1}}
\cortext[cor1]{Corresponding author.}
\ead{robertej@mit.edu}

\author[label1]{J.~Bernauer\fnref{c}}
 \fntext[c]{Present Address: Department of Physics and Astronomy,
Stony Brook University, Stony Brook, NY 11794, USA}
\author[label3]{C.~M.~Cooke}
\author[label1]{R.~Corliss}
\author[label1]{C.~S.~Epstein}
\author[label1]{P.~Fisher}
\author[label1]{I.~Fri\v{s}\v{c}i\'{c}}
\author[label1]{D.~Hasell}
\author[label2]{E.~Ihloff}
\author[label2]{J.~Kelsey}
\author[label1]{S.~Lee}
\author[label1]{R.~G.~Milner}
\author[label1]{P.~Moran}
%%\ead{cepstein@mit.edu}
\author[label1]{S.~G.~Steadman}
\author[label2]{C.~Vidal}

\address[label1]{Laboratory for Nuclear Science, Massachusetts Institute of Technology, Cambridge, MA 02139}
\address[label2]{MIT Bates Research \& Engineering Center, Middleton, MA 01949}
\address[label3]{High Voltage Research Laboratory, Research Laboratory for Electronics, Massachusetts Institute of Technology, Cambridge, MA 02139 }

\begin{abstract}
%% Text of abstract

The design, construction, installation, and testing of a Faraday Cup intended to measure the current of a 3~MeV, \SI{1}{\micro\ampere}  electron beam is described. Built as a current monitor for a \Moller scattering measurement at the MIT High Voltage Research Laboratory, the device combines a large angular acceptance with the capability to measure a continuous, low energy beam. Bench studies of its performance demonstrate current measurements accurate to the percent level at \SI{1}{\micro\ampere}. The Faraday Cup was designed and constructed at MIT and has been in use at the HVRL since 2017,  providing a significantly more detailed measurement of beam current than was previously available.
\end{abstract}


\end{frontmatter}

\section{Introduction}
Faraday Cups are regularly used to measure beam current in electron and ion beam experiments. For a detailed description of basic Faraday Cup principles, see \cite{generalxx}. A wide range of sizes and types have been constructed for a variety of purposes ranging from a simple determination of beam current, to obtaining a 2D beam profile \cite{Density}, as well as measuring the current of laser-generated plasma particles \cite{laser}, and of high-intensity pulsed electron beams \cite{pulses}. 

The Faraday Cup described in this article was designed to measure the current of a 3\,MeV continuous electron beam from a \VDG accelerator at MIT's High Voltage Research Laboratory (HVRL). This current measurement is an integral part of the \Moller scattering experiment at the HVRL, which measures \Moller scattering cross-sections in a low energy regime where the electron mass is not negligible. 

It is important to quantify radiative \Moller scattering as it has not yet been measured precisely at these energies \cite{Epstein2016} and is an experimental concern for other proposed experiments \cite{DL2018}. Recently, the experiment has measured the radiative \Moller electron momentum spectrum at three different spectrometer angles between 30\degrees and 40\degrees \cite{DL2018}. The current measurement from the Faraday Cup allows for the normalization of rates between different spectrometer positions and beam currents, and is necessary to reconstruct an absolute cross-section. 

\subsection{Design Considerations}

\begin{figure}[H]
\centering				
\includegraphics[width=\linewidth]{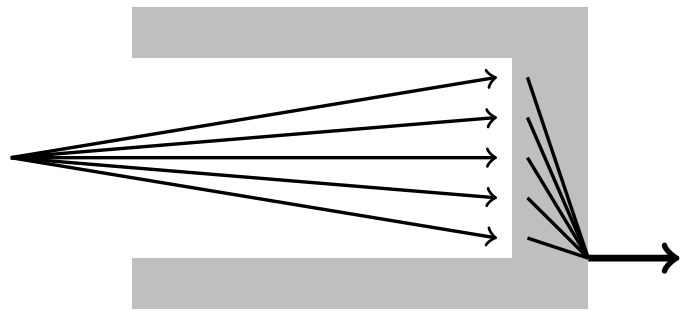}
\caption{Functional sketch of a Faraday Cup. A charged beam (black arrows) collects on a cup which is electrically connected to an ammeter or other measuring device. The electrons from the beam flow through the cup out to the measuring device. The cup must be designed appropriately to stop the beam and capture all charge.}
\label{cup}
\end{figure}

In its simplest form, the device consists of one or more conducting materials formed into the shape of a cylinder with an inner cavity, giving the approximate shape of a cup (\Fig{cup}). Incoming electrons (or ions) from a beam strike the conducting material and are absorbed.  The charge then flows to ground through an electrometer, from which the beam current can be determined. 

An ideal Faraday Cup would perfectly measure the beam current: it would be physically large enough to have all electrons from the beam pass into the cup, have thick enough walls and appropriate other features to fully capture the charge of each electron, and be designed such that all net charge would flow to ground through the ammeter. In designing this device, we considered three deviations from ideal behavior: electrons back-scattering out of the cup, electrons escaping through the cup due to insufficiently thick walls, and current leaking to ground, bypassing the ammeter. More-exotic beamlines have commensurately complex issues, such as measuring a low average current on a high intensity pulsed beam, but these are not relevant for the low-energy, continuous-wave (CW) electron beam for which this Faraday Cup was designed.

A combination of conductive materials with varying densities can be used to ensure the capture of all electrons, while satisfying size constraints. Low-Z materials are often chosen for the surface of the cup to minimize backscatter, with more dense material added behind to ensure absorption.
To further minimize back-scattering, electric and magnetic fields can be added to the front of the Faraday Cup to help contain the electrons once they enter the Cup. To mitigate leakage currents to ground, Faraday Cups are usually placed into the vacuum system of the beamline, and therefore require a non-conductive support structure and a vacuum-compatible construction.

\section{Design and Construction of the Faraday Cup}
\subsection{Mechanical Design and Construction}

This Faraday Cup was designed specifically for the beam conditions at the HVRL: a low intensity electron beam operated at most with a micro-ampere of current. Electrons from this beam have a range of approximately \SI{1}{cm} in graphite \cite{NIST2015}, which is small enough that a single-material absorber is sufficient to stop the electrons. The power deposited by this \SI{1}{\micro\ampere} beam is 3\,W, which is easily dissipated and produces a negligible amount of heating.

The opening diameter of the graphite cup is constrained by the size of the beam at the entrance to the Faraday Cup. In this particular experiment, the electron beam is incident on a \SI{2}{\micro\meter} diamond-like carbon target. Using a Gaussian approximation for the spread of the beam due to multiple scattering \cite{PDG}, this introduces approximately 0.4\degrees of angular spread, which corresponds to an RMS beam diameter of approximately \SI{1.5}{\cm} at the entrance to the Faraday Cup. The inner graphite cup was thus designed to be roughly three times larger than this, corresponding to collecting greater than 99\% of all electrons, using this RMS beam estimation. Using a stock 3'' diameter graphite rod, a cup with an opening diameter of approximately \SI{4.6}{\cm} was machined, which satisfies this diameter requirement while also keeping the cup walls thicker than the electron range. 
 
\begin{figure}[htb]
\centering				
\includegraphics[width=\linewidth]{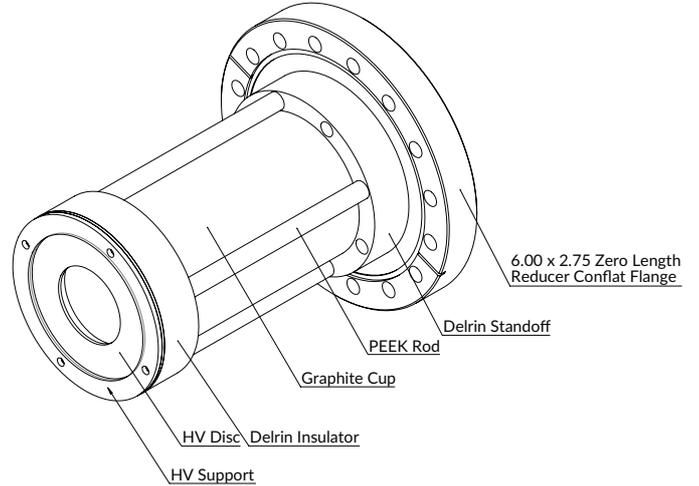}
\caption{Design of the Faraday Cup: a graphite cup is supported by a Delrin and PEEK structure, which is mounted on a stainless steel Conflat flange. The aluminum HV bias disc is attached to the support structure at the front of the Faraday Cup. Not shown in the figure are insulated wires connecting both the HV bias disc and the graphite cup to their respective feedthroughs.}
\label{Assem1}
\end{figure}

\begin{figure}[htb]
\centering				
\includegraphics[width=\linewidth]{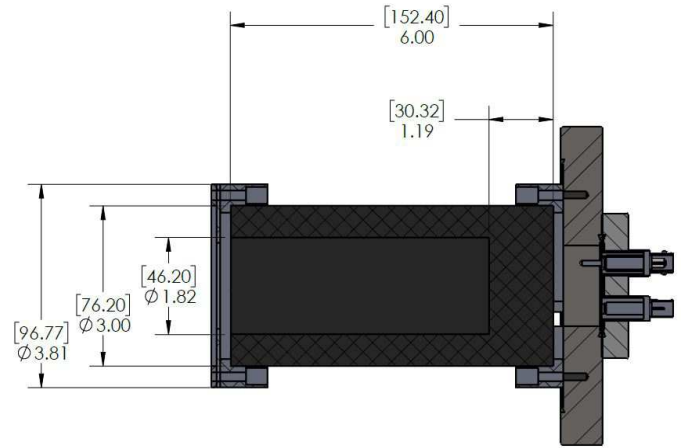}
\caption{Faraday Cup design with dimensions labeled. Bracketed dimensions are in millimeters, unbracketed in inches.}
\label{Assem2}
\end{figure}

\begin{figure}[htb]
\centering
\begin{subfigure}{0.5\linewidth}
\centering
\includegraphics[width=.75\linewidth,angle=0]{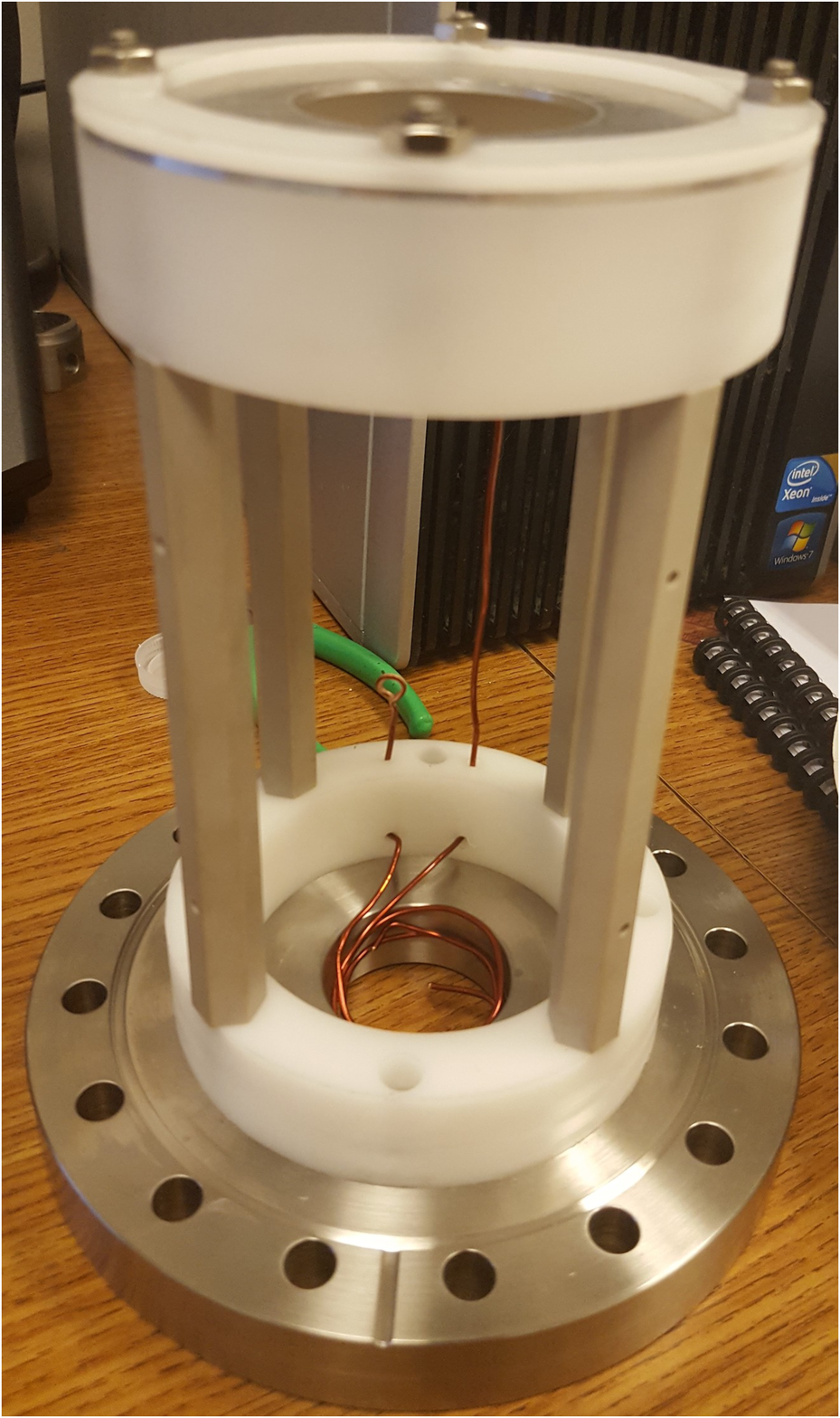}
%\caption{A subfigure}
\label{basewithrods}
\end{subfigure}%
\begin{subfigure}{0.5\linewidth}
\centering
\includegraphics[width=.74\linewidth,angle=0]{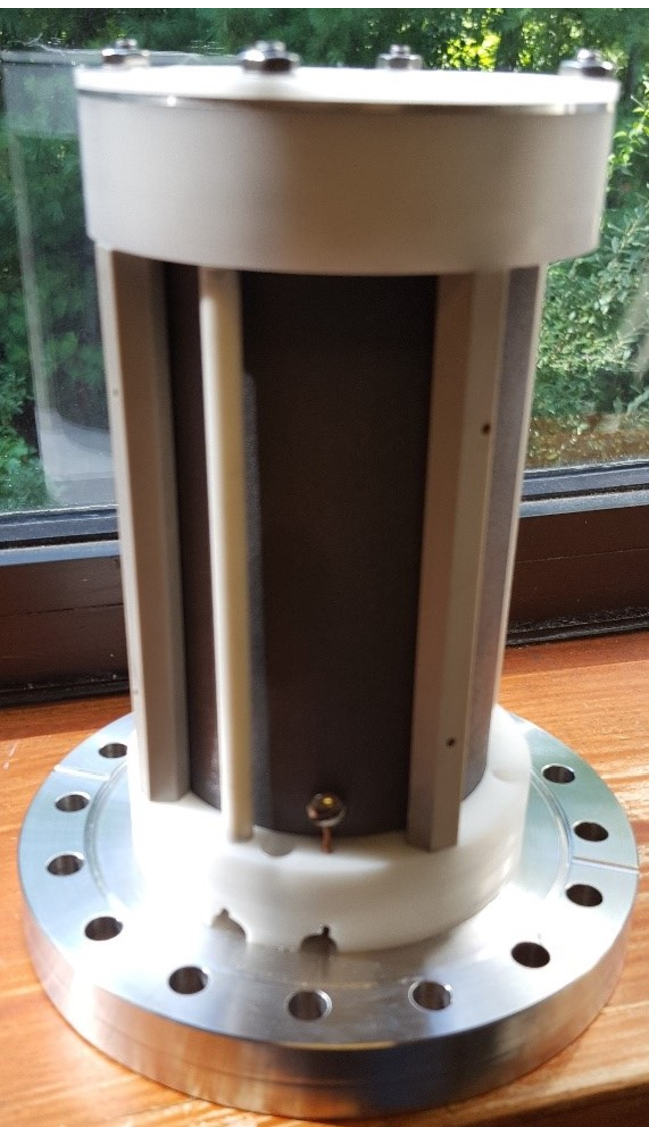}
%\caption{A subfigure}
\label{finishedcup}
\end{subfigure}
\caption{Left: Faraday Cup support structure with Conflat flange at base, Delrin supports, and PEEK rods. The two visible insulated wires attach to the graphite cup and to the high voltage biasing disc. Right: Completed Faraday Cup with graphite cup secured in support structure and electrical connecting screw attached at its base.}
\label{finishedpics}
\end{figure}

The inner conducting cup requires an insulating support structure and vacuum housing with electrical feedthroughs. Because of the cup's relatively light weight, a simple plastic standoff structure was sufficient to hold it in place. This structure was machined out of Delrin, with supporting rods constructed from PEEK. These supporting standoffs were mounted on a vacuum flange attached to the end of the beamline. A thin, high voltage biasing disc designed to hold up to \SI{500}{V} was also included in the design, to aid in trapping back-scattered electrons. This disc overhangs the inner opening of the graphite cup and has an inner diameter of \SI{4.1}{\cm}. The Faraday Cup was constructed at the MIT Bates Research and Engineering Center. A design drawing is shown in \Fig{Assem1}, and the specifications for the graphite cup are given in \Fig{Assem2}. A photo of the assembled cup is shown in \Fig{finishedpics}.

\subsection{Electrical Readout and Circuit Design}

The Faraday Cup was connected to a Keithley 6517b Electrometer in order to precisely measure the beam current. This device is capable of making high-precision current measurements in the range of \SI{20}{\pico\ampere} to \SI{20}{\milli\ampere} with low input voltage burden \cite{Keithley}. More complex experiments with bunched beams sometimes require an array of resistors and capacitors to smooth the readout current, but this is not necessary with the CW beam at the HVRL. The simple readout design of this cup makes it sensitive to higher-frequency fluctuations in the beam current, which can be informative in the diagnosis of beam-related abnormalities.

%$^{14}$

\begin{figure}[H]
\centering				
\includegraphics[width=0.64\linewidth]{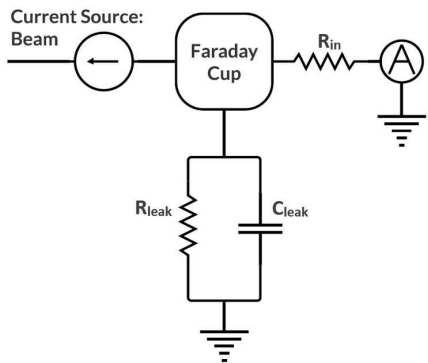}
\caption{Circuit Schematic for Faraday Cup. }%Note that the \SI{500}{V} bias disc is electrically separate from the Faraday Cup but is physically close.}
\label{schematicAAA}
\end{figure}

Fig. \ref{schematicAAA} shows the equivalent electrical circuit for this design. Current, in the form of the electron beam, flows into the Faraday Cup. The Faraday Cup has a finite but very large  %(>\SI{10}{\ohm})
leakage resistance to ground. As this resistance is many orders of magnitude larger than the resistance to the ammeter, it is not necessary to measure it precisely, although it is estimated to be above $10^{15}$ \SI{}{\ohm} based off the Faraday Cup design and material resistivities. The Faraday Cup also has capacitance to ground, which is estimated to be on the order of \SI{10}{\pico\farad} based on geometric considerations. The current from the Faraday Cup flows through an RG-58 cable to the electrometer where it is measured and read out at approximately 10 Hz. %. The electrometer is read out in a LabView interface at approximately 10 Hz.%connected to a computer via a direct RS-232 connection, and read out in a LabView interface. 

\section{Testing and Commissioning}

After the completion of construction and cleaning, the Faraday Cup was connected to a testing chamber for leak-checking, and then tested for electrical integrity.

\subsection{Electrical Testing}

\begin{figure}[H]
\centering				
\includegraphics[width=\linewidth]{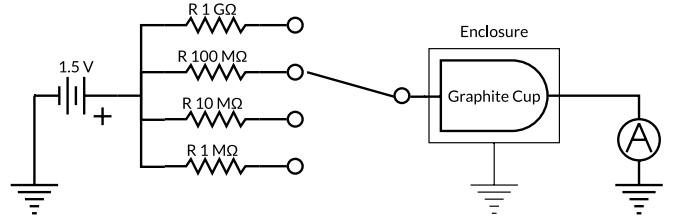}
\caption{Circuit Schematic of Faraday Cup test setup.}
\label{schematic}
\end{figure}

The Faraday Cup and electrometer were both tested using a \SI{1.5}{V} nominal AAA battery with four 1\% resistors (\SI{1}{\mega\ohm} to \SI{1}{\giga\ohm}). The battery and resistors were connected to a BNC termination, providing a current source (Fig. \ref{schematic}).% and the testing layout in Fig. \ref{layout}.

\begin{figure}[H]
\centering				
\includegraphics[width=0.96\linewidth]{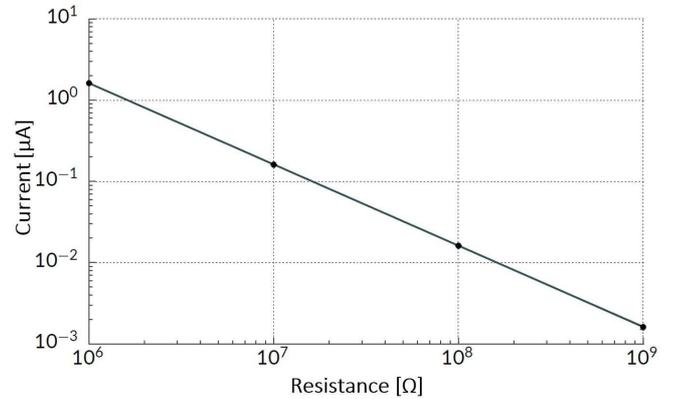}
\caption{Current versus resistance through the Faraday Cup.}
\label{current}
\end{figure}

As can be seen in \Fig{current}, the response of the Faraday Cup to the current is very linear, indicative of proper electrical connections throughout the device and of sufficiently high insulation resistance. The measured current in the ammeter corresponds very closely to the calculated supplied current, lying well within the 1\% uncertainty originating from the circuit resistor. These results combined give confidence that the Faraday Cup was constructed and assembled properly and there were no anomalous electrical effects.

\subsection{Beam Testing at the HVRL}

\begin{figure}[htb]
\centering				
\includegraphics[width=0.5\linewidth]{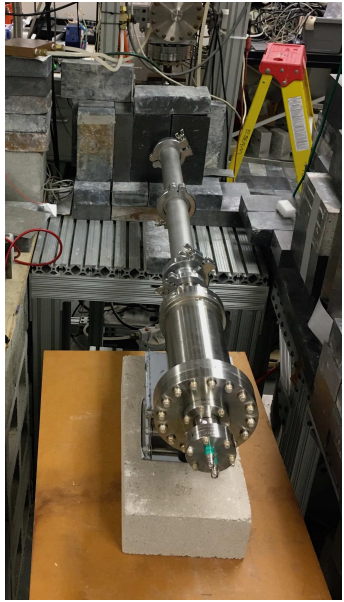}
\caption{Faraday Cup connected to the beamline at the HVRL.}
\label{hvrl}
\end{figure}

Upon completion of studies, the Faraday Cup was attached to the beamline, and was tested directly with the electron beam (\Fig{hvrl}). While we do not have an absolute method to compare the current measured with the Faraday Cup to that of a precisely-known source, the measured current is consistent with that expected from previous operating experience.

\begin{figure}[H]
\centering				
\includegraphics[width=\linewidth]{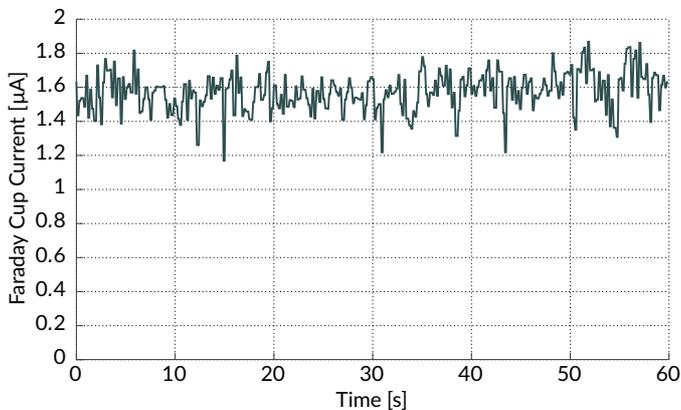}
\caption{Time structure of the HVRL beam as shown in a one minute sample from the November 2017 trial run.}
\label{data}
\end{figure}

Data from the Faraday Cup is shown in \Fig{data}. A notable feature is that there appears to be oscillatory behavior: it is assumed that this reflects real variation in the beam current due to the electrical architecture of the Van de Graaff accelerator. It is also a possible sign of unintentional oscillations in the readout circuit coupling to the accelerator electrical system, although this is less likely as the readout circuit has been tested independently and no such oscillations were found.  

There was no observable difference in the measured current with and without the \HVBD energized. This is expected, as the \HVBD can be estimated to change the accuracy of the Faraday Cup at the percent level based off of reports from other Faraday Cup designs (\cite{generalxx}), while the beam current fluctuations are on the order of 10\%.The magnitude of these beam fluctuations were previously unknown, as the facility formerly only had current-measuring capabilities accurate to an order of magnitude. We are now able to investigate the source of these fluctuations and monitor the improvement of the beam with this device.

\section{Discussion and Future Directions}

This Faraday Cup marks a significant improvement over previous beam current diagnostics at the HVRL. This has allowed a recent experiment to take place and extract \Moller cross-sections at low energy. To increase the precision of the experiment and enable an absolute current measurement, we will require a better understanding of the beam-target interaction and how it affects the Faraday Cup acceptance. This can be performed, for example, by systematically varying target thickness or beam position. These measurements are currently limited by the long- and short-term stability of the beam. However, upgrades are under consideration that would improve the stability of the beam, allowing for improved accuracy in the current measurement. The Faraday Cup will be able to confirm that these upgrades are effective, and will enable higher precision absolute measurements in the future.

\section*{Acknowledgements}

This research is supported by the U.S. Department of Energy Office of Nuclear Physics (Grant No. DE-FG02-94ER40818), the U.S. Department of Energy Office of High Energy Physics (Grant No. DE-SC0011970), and the NSF MRI Program (Award No. 1437402).

\bibliographystyle{elsarticle-num}
\bibliography{pd2}

\end{document}